\documentclass[10pt,notitlepage]{elsarticle}
\usepackage[utf8]{inputenc}
\usepackage{amsmath}
\usepackage{amsfonts}
\usepackage{amssymb}
\usepackage{graphicx}
\usepackage{textcomp}
\usepackage{enumitem}
\setlist{  
  listparindent=\parindent,
  parsep=0pt,
}
\usepackage{natbib}
\usepackage{tikz}
\usetikzlibrary{decorations.pathreplacing}
\usepackage{color}
\usepackage[figuresright]{rotating}
\usepackage[left=1in,right=1in,top=1in,bottom=1in]{geometry}

\begin{document}
\author[mit]{Jayson R.~Vavrek\corref{cor1}}
\ead{jvavrek@mit.edu}
\author[mit]{Brian S.~Henderson}
\ead{bhender1@mit.edu}
\author[mit]{Areg Danagoulian}
\ead{aregjan@mit.edu}
\title{High-accuracy Geant4 simulation and semi-analytical modeling of nuclear resonance fluorescence}
\cortext[cor1]{Corresponding author; tel 1-617-324-4678, fax 1-617-253-5405}
\address[mit]{Laboratory for Nuclear Security and Policy, Massachusetts Institute of Technology, 77 Massachusetts Avenue, Cambridge, MA, USA, 02139}

\begin{abstract}
Nuclear resonance fluorescence (NRF) is a photonuclear interaction that enables highly isotope-specific measurements in both pure and applied physics scenarios. High-accuracy design and analysis of NRF measurements in complex geometries is aided by Monte Carlo simulations of photon physics and transport, motivating Jordan and Warren (2007) to develop the G4NRF codebase for NRF simulation in Geant4. In this work, we enhance the physics accuracy of the G4NRF code and perform improved benchmarking simulations. The NRF cross section calculation in G4NRF, previously a Gaussian approximation, has been replaced with a full numerical integration for improved accuracy in thick-target scenarios. A high-accuracy semi-analytical model of expected NRF count rates in a typical NRF measurement is then constructed and compared against G4NRF simulations for both simple homogeneous and more complex heterogeneous geometries. Agreement between rates predicted by the semi-analytical model and G4NRF simulation is found at a level of ${\sim}1\%$ in simple test cases and ${\sim}3\%$ in more realistic scenarios, improving upon the ${\sim}20\%$ level of the initial benchmarking study and establishing a highly-accurate NRF framework for Geant4.
\end{abstract}

\begin{keyword}
nuclear resonance fluorescence \sep G4NRF \sep Geant4 \sep benchmarking \sep verification
\end{keyword}

\maketitle


\section{Introduction}
In recent years, nuclear resonance fluorescence (NRF)---the resonant absorption and re-emission of photons by a nucleus---has been widely-proposed as a powerful isotope-specific assay technique. Nuclear weapon treaty verification~\cite{vavrek2018experimental,kemp2016physical}, spent fuel measurement~\cite{quiter2011transmission}, and cargo scanning~\cite{bertozzi2005nuclear, pruet2006detecting} systems use NRF as an active interrogation technique to discern the isotopics of or detect the presence and quantity of special nuclear materials. In the domain of pure physics, NRF is useful as a probe of nuclear structure across a broad array of isotopes~\cite{kneissl1996investigation}.

For realistic experimental geometries, expected NRF count rates may be calculated through Monte Carlo simulation of photon and electron transport and physics. The G4NRF~\cite{jordan2007simulation} package for the Geant4~\cite{agostinelli2003geant4} Monte Carlo toolkit was developed by Jordan and Warren at Pacific Northwest National Laboratory, while NRF data libraries for MCNPX~\cite{pelowitz2011mcnpx} have been developed by Wilcox et al.~at Los Alamos National Laboratory~\cite{wilcox2011mcnpxnrf}. In the former case, NRF rates predicted by the G4NRF code were initially only benchmarked against theory to within ${\sim}20\%$ and validated against data to within a factor of ${\sim}3$~\cite{warren_pers_comm}. Moreover, the initial study made a number of mathematical simplifications in its analytical model: it neglected non-resonant photon attenuation (e.g., Compton scattering), assumed the emission of NRF photons to be isotropic, and implemented a Gaussian approximation to the NRF cross section that is not valid for thick targets or large resonance widths. This benchmarking study accounts for these three effects and therefore presents improved benchmarking of the G4NRF code against a more accurate semi-analytical radiation transport model. To this end, Section~\ref{sec:model} first constructs this high-accuracy semi-analytical model for the NRF photon count rate observed by a detector. A series of Geant4+G4NRF Monte Carlo simulations is then compared against the semi-analytical model in Section~\ref{sec:verification}. Section~\ref{sec:discussion} concludes with a discussion of results.

\section{Semi-analytical model for NRF count rates}\label{sec:model}
In this section we present a model for predicting the absolute NRF count rate observed by a detector in a transmission NRF measurement. The model is based upon the NRF cross section and radiation transport development previously given in the literature (primarily Refs.~\cite{metzger1959resonance, quiter2010thesis}), but expands the treatment to multiple-isotope targets and practical considerations for high-accuracy computation. First, we derive the NRF cross section necessary for both the semi-analytical and G4NRF rate predictions. We will then show that the shape of the NRF cross section can influence NRF count rate predictions substantially for thick targets, motivating the derivation and use of highly accurate cross section formulae. We then apply the NRF cross section to the radiation transport problem that describes a generic transmission NRF measurement in order to construct a semi-analytical model for the expected NRF count rate. Such semi-analytical calculations necessarily involve approximations to keep the mathematics tractable, and thus are somewhat limited in the experimental complexity they can accurately model. However, they offer a powerful tool for investigating the dependence of NRF count rates on various physics or geometrical parameters without running computationally expensive simulations, and are useful in verifying the implementation and accuracy of the G4NRF code.

\subsection{NRF cross sections}\label{sec:cross_section}
Nuclear resonance fluorescence (NRF) describes the X$(\gamma,\gamma')$X reaction in which a nucleus X with resonance energy $E_r$ resonantly absorbs a photon of energy $E\simeq E_r$, thereby transitioning from its ground state to the excited state at $E_r$ \cite{kneissl1996investigation,metzger1959resonance}. The excited nucleus subsequently decays after time $\mathcal{O}(\text{fs})$, re-emitting a photon of energy $E' \simeq E$ (neglecting the relatively small nuclear recoil given later in Eq.~\ref{eq:E_rec}) if the decay is direct to the ground state, or a lower energy $E' \simeq E-E_j$ if the decay proceeds first through an intermediate level $j$.

The NRF cross section (at temperature $T=0$~K) for absorption through an isolated resonance at energy level $E_r$ followed by decay to an energy level $E_j$ may be found (using, e.g., perturbation theory~\cite{bethe1937resonance}) to follow a single-level Breit-Wigner profile~\cite{metzger1959resonance}:
\begin{align}\label{eq:sigmaBWj}
\sigma_{r,j;\text{\,0\,K}}^\text{NRF}(E) = \pi g_r \left( \frac{\hbar c}{E} \right)^2 \frac{\Gamma_{r,0} \Gamma_{r,j}}{(E-E_r)^2 + (\Gamma_r/2)^2}.
\end{align}
The $g_r$ term is a statistical factor arising from the number of available nuclear spin and photon polarization states, given by
\begin{align}
g_r = \frac{2J_r+1}{2(2J_0+1)},
\end{align}
where $J_0$ and $J_r$ are the ground-state and resonant-level nuclear spins, respectively. The $\Gamma_{r,0}$ and $\Gamma_{r,j}$ terms denote the partial widths for decay from the level at $E_r$ to $E_j$, while the $\Gamma_r$ is the total width of the level, i.e., the sum of the partial widths. For most calculations in this work, it is more convenient to work with the cross section for absorption only,
\begin{align}\label{eq:sigmaBWabs}
\sigma_{r;\text{\,0\,K}}^\text{NRF}(E) = \sum_j \sigma_{r,j;\text{\,0\,K}}^\text{NRF}(E) = \pi g_r \left( \frac{\hbar c}{E} \right)^2 \frac{\Gamma_{r,0} \Gamma_{r}}{(E-E_r)^2 + (\Gamma_r/2)^2},
\end{align}
which differs from the absorption+decay cross section only by the level's branching ratio $b_{r,j}\equiv \Gamma_{r,j} / \Gamma_{r}$, with the normalization condition $\sum_j \Gamma_{r,j} = \Gamma_r$, i.e., $\sum_j b_{r,j} = 1$.

At non-zero temperatures, the NRF absorption cross section is most accurately described by a Doppler-broadened version of the Breit-Wigner distribution in Eq.~\ref{eq:sigmaBWabs}:
\begin{align}\label{eq:sigmaNRF}
\sigma_r^\text{NRF}(E) = 2\pi^{1/2} g_r \left( \frac{\hbar c}{E_r} \right)^2 \frac{b_{r,0}}{\sqrt{t}} \int_{-\infty}^{+\infty} \exp\left[ -\frac{(x-y)^2}{4t} \right] \frac{dy}{1+y^2},
\end{align}
which integrates over the thermal distribution of speeds of the target nuclei~\cite{metzger1959resonance}. Here we have suppressed the temperature subscript for brevity, replaced the $1/E^2$ term with $1/E_r^2$ (valid near the resonance), and defined
\begin{align}
x &\equiv 2(E-E_r)/\Gamma_r,\\
t &\equiv (\Delta/\Gamma_r)^2,
\end{align}
where $\Delta$ is the width of the level after Doppler broadening, with
\begin{align}\label{eq:DopplerDelta}
\Delta = E \sqrt{\frac{2kT}{Mc^2}},
\end{align}
where $k$ is Boltzmann's constant, $T$ is the absolute temperature, and $Mc^2$ is the rest-mass energy of the nucleus.\footnote{For greater accuracy, the temperature $T$ may be replaced by the `effective' temperature~\cite{metzger1959resonance}
\begin{align}
T_\mathrm{eff} = 3 T \left( \frac{T}{\theta_D} \right)^3 \int_0^{\theta_D/T} x^3 \left( \frac{1}{e^x -1} + \frac{1}{2} \right) dx,
\end{align}
where the absorber's Debye temperature $\theta_D$ accounts for the effect of the atomic lattice on the ideal Maxwell-Boltzmann distribution of speeds. This change propagates through to the $\Delta$ of Eq.~\ref{eq:DopplerDelta} and thus the $t$ of Eq.~\ref{eq:sigmaNRF}. G4NRF uses $T_\text{eff}$ if $\theta_D$ is known, and defaults to $T=300$~K otherwise.} The NRF cross section given by Eq.~\ref{eq:sigmaNRF} is implemented in both the semi-analytical NRF rate model in the next section and the G4NRF Monte Carlo code.

A useful measure of the `strength' of a resonance is the \textit{integrated cross section},
\begin{align}\label{eq:sigma_int}
\int \sigma_r^\text{NRF}(E)\,dE = 2\pi^2 g_r \left( \frac{\hbar c}{E_r} \right)^2 \Gamma_{r,0},
\end{align}
which can be found by approximating $(\hbar c /E)^2 \simeq (\hbar c /E_r)^2$ and $E_r \gg \Gamma_r$ then integrating Eq.~\ref{eq:sigmaBWabs} over $E \in [0,\infty)$, or by approximating $\Delta \simeq E_r \sqrt{2kT/Mc^2} \gg \Gamma_r$ then integrating Eq.~\ref{eq:sigmaNRF} over $x \in (-\infty, \infty)$. As will be shown later, the expected NRF count rate in an experiment is proportional to the integrated cross section in the thin-target limit.

The fundamental parameters required in Eq.~\ref{eq:sigmaNRF}---and thus determined for each resonant level in an NRF cross section measurement---are the level's width $\Gamma_r$ and set of branching ratios $b_{r,j}$. The NRF transitions studied in this work are all transitions directly to the ground state, such that only the branching ratios to the ground state $b_{r,0}$ are necessary. Cross section parameters reported by different experiments (or tabulated in the ENSDF databases, e.g., Ref.~\cite{nndc2015u238}) can vary drastically, however. The ground-state branching ratio of the U-238 2.245~MeV level, e.g., differs by ${\sim}30\%$ between ENSDF and Ref.~\cite{quiter2011transmission}, while its width $\Gamma_r$ differs by more than an order of magnitude. Similarly, the values of $\Gamma_{r,0}^2/\Gamma_r$ reported by Ref.~\cite{quiter2011transmission} and Ref.~\cite{hammond2012dipole} differ by ${\sim}25\%$ in the U-238 $2.209$~MeV line, and by a factor of $6$ in the U-238 $2.468$~MeV line. These discrepancies may introduce systematic uncertainties much larger than our desired accuracy for the verification study; for consistency, then, both the calculations and simulations in this work use an assumed set of cross section parameters (in isotopes relevant to nuclear security applications---see Refs.~\cite{vavrek2018experimental,kemp2016physical}) from various references as shown in Table~\ref{tab:xsec_params}. For the U-238 resonances, preference is given to experimentally-determined (i.e., not ENSDF-evaluated) data; specifically, the integrated cross sections and ratios of widths in Table~1 of Ref.~\cite{quiter2011transmission} (which derive from Ref.~\cite{heil1988observation}) are used to infer values of $\Gamma_r$ and $b_{r,0}$ for the three major U-238 resonances. No NRF data on Pu-240 exists in ENSDF, so values of $\Gamma_r$ and $b_{r,0}$ are similarly inferred from experimental data in Table~II of Ref.~\cite{quiter2012nrf}. For the $2.212$~MeV resonance of Al-27, the value of $\Gamma_r$ is determined from the lifetime listed in Table~27.4 of Ref.~\cite{endt1990levels} and $b_{r,0}$ is found using Table~27.6 of the same work. We note that while some of the U-238 cross section parameters may vary significantly across references, the Al-27 parameters generally agree to within a few percent. Since the $\Gamma_r$ and $b_{r,0}$ values read by G4NRF are stored in plaintext files, the user may configure G4NRF to use a custom set of cross section parameters.

\begin{table}[!h]
\centering
\begin{tabular}{c|c|c|c|c|c}
isotope & $E_r$ [MeV] & $\Gamma_r$ [meV] & $b_{r,0}$ & $\int \sigma_r^\text{NRF}(E)\, dE$ [eV$\cdot$b] & Ref.\\\hline
Al-27 & 2.212 & 17.1 & 0.997 & 18.0 & \cite[Tables 27.4, 27.6]{endt1990levels} \\
U-238 & 2.176 & 54.7 & 0.658 & 87.7 & \cite[Table 1]{quiter2011transmission} \\
U-238 & 2.209 & 54.3 & 0.645 & 82.7 & \cite[Table 1]{quiter2011transmission} \\
U-238 & 2.245 & 28.9 & 0.680 & 45.0 & \cite[Table 1]{quiter2011transmission} \\
Pu-240 & 2.433 & 81.3 & 0.655 & 103.7 & \cite[Table II]{quiter2012nrf} \\
\end{tabular}
\caption{Assumed NRF cross section parameters used in this work. The integrated cross section $\int \sigma_r^\text{NRF}(E)\, dE$ is directly proportional to the product $\Gamma_r\,b_{r,0} \equiv \Gamma_{r,0}$ as shown in Eq.~\ref{eq:sigma_int}.}
\label{tab:xsec_params}
\end{table}

Due to conservation of energy and momentum, a free nucleus undergoing NRF will recoil with kinetic energy $E_\text{rec}$ determined by the Compton-like formula
\begin{align}\label{eq:E_rec}
E_\text{rec}= E \left[ 1- \frac{1}{1+E(1-\cos\chi)/Mc^2} \right] \simeq \frac{E^2}{Mc^2}(1-\cos\chi),
\end{align}
where $\chi$ is the photon scattering angle relative to its incident direction, and $E \ll Mc^2$ has been applied in the Taylor expansion. For nuclei bound in an atomic lattice, $E_\text{rec}$ may be large enough to overcome the lattice displacement energy $E_d$ (${\gtrsim}10$~eV in pure metals~\cite{astme521}) in which case the kinetic energy transfer is $E_\text{rec}-E_d$. If the value of $E_\text{rec}$ for an unbound nucleus is less than $E_d$, the recoil is instead transferred to the entire lattice: $M\to\infty$, $E_\text{rec}\to 0$, and recoilless NRF (the M\"ossbauer effect) is achieved~\cite{mossbauer1958nrf}. Values of $E_\text{rec}$ vary from 0 in the forward ($\chi=0$) direction to $2E^2/Mc^2$ in the backward ($\chi=\pi$) direction, and average to $E^2/Mc^2$ over all solid angles. For a photon energy of $E=2.2$~MeV incident on U-238 or Al-27 and re-emitted at the $\chi = 125^\circ$ commonly used in experiments, $E_\text{rec} \simeq 35$~eV or $300$~eV, respectively. Since the outgoing photon energy is correspondingly reduced by $E_\text{rec} \gg \Delta$, photons emitted in the backwards direction (which produce relatively large nuclear recoils) will be well below the resonance energy $E_r$ and highly unlikely to undergo another NRF interaction. This lack of self-attenuation in the backwards direction, along with high backgrounds in the forward direction, gives rise to experiments in which NRF photon spectra are measured in backwards emission directions, as discussed in the following section.

\subsection{NRF count rates in a transmission measurement}

\begin{figure}[h!]
\centering
\includegraphics[width=0.85\columnwidth]{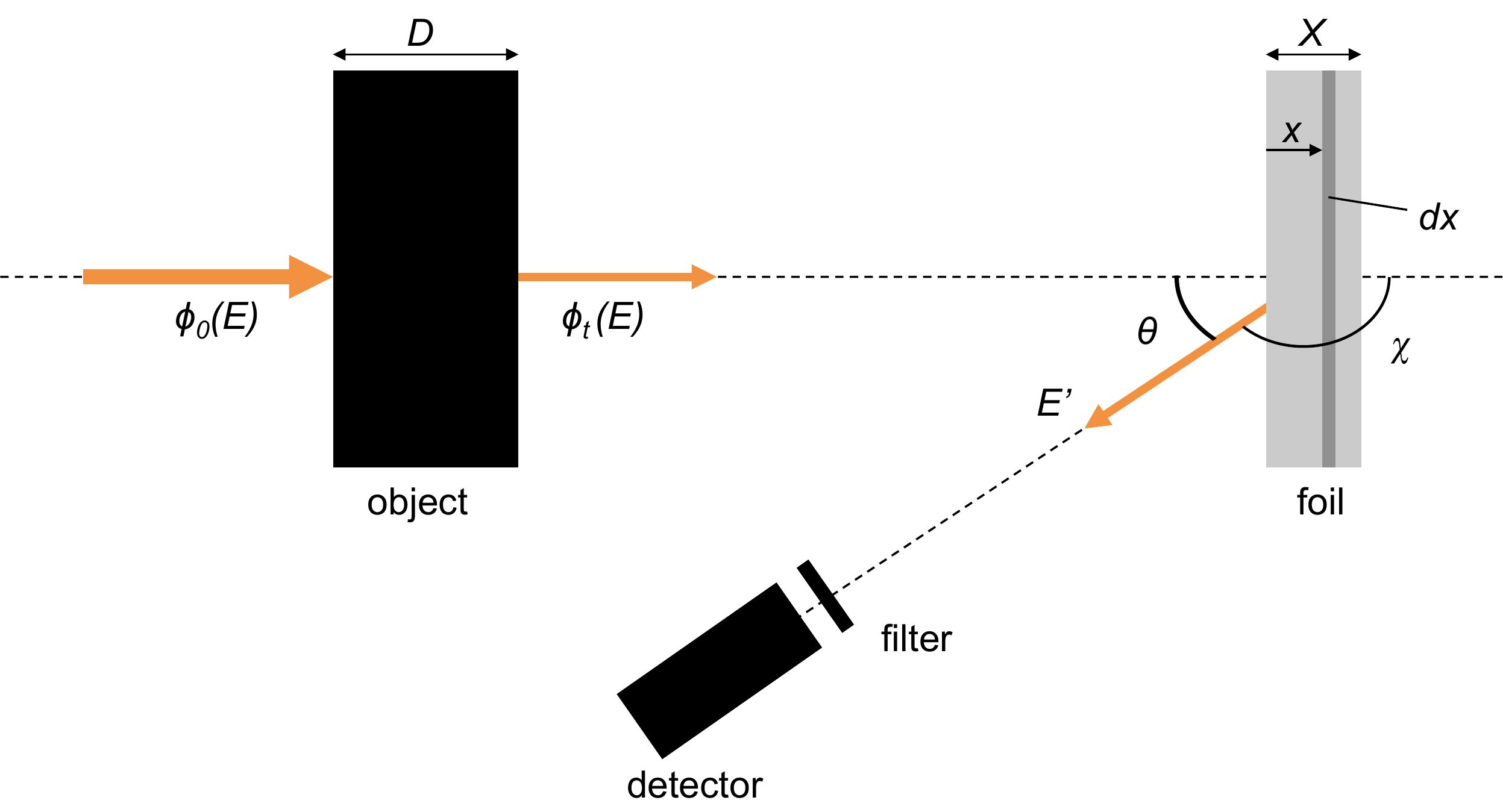}
\caption{Schematic of a transmission NRF measurement. In this 1D, single-isotope model, a narrow parallel beam of the interrogating flux $\phi_0(E)$ impinges on the target (of linear thickness $D$) to be measured, and the flux $\phi_t(E)$ transmitted through the target impinges on a reference foil (of linear thickness $X$) composed of the same isotope as the target.}
\label{fig:schematic}
\end{figure}

In this section, we use the NRF cross section (Eq.~\ref{eq:sigmaNRF}) to construct an expression for the expected NRF count rates in a transmission NRF measurement, replicating the treatment given in Ref.~\cite{quiter2010thesis}. Consider the transmission NRF experimental setup in Fig.~\ref{fig:schematic}, in which a parallel beam of incident flux $\phi_0(E)$ impinges on the object to be measured. Resonant (NRF) and non-resonant (Compton scattering, pair production, photoelectric absorption\footnote{In reality there is a small non-resonant contribution from coherent scattering processes in which the photon undergoes a change in direction but not energy. As of Geant4 v10.3, coherent scattering processes are not included in the standard electromagnetic libraries and thus are excluded here.}) interactions in the measurement object attenuate the incident flux $\phi_0(E)$, producing a transmitted flux
\begin{align}\label{eq:phi_t}
\phi_t(E)=\phi_0(E) \exp(-D[\mu_\mathrm{NRF}(E) + \mu_\mathrm{nr}(E)]),
\end{align}
where $\mu_\text{NRF}(E) \equiv N\sigma_r^\text{NRF}(E)$ denotes the linear attenuation coefficient for a target number density $N=\rho N_\text{Av}/A$ and NRF cross section for a single resonance (Eq.~\ref{eq:sigmaNRF}, suppressing the subscript $r$), and $\mu_\text{nr}$ denotes non-resonant attenuation (as given by, e.g., the NIST XCOM databases~\cite{nist_xcom}).
Since $\mu_\mathrm{NRF}(E)$ provides strong attenuation over a narrow (${\sim}\Delta$) energy range, the transmitted flux is preferentially attenuated (`notched') at the resonant energy $E_r$, and the depth of the notch encodes information about the areal density $\rho D$ of the measurement object. Direct measurement of the notch is impractical, however, since detector energy resolution is typically much larger than the thermal width of the notch $\Delta$, and irreducibly high backgrounds are produced in the forward direction. An indirect approach is taken instead: the notched flux $\phi_t(E)$ impinges on a reference foil constructed of the NRF isotope(s) of interest, so that the remaining photons near the resonant energy $E_r$ undergo NRF in the reference foil (see again Fig.~\ref{fig:schematic}). The resulting NRF flux in the \textit{backwards} direction then produces a triple-differential rate of NRF detections\footnote{Note here that we are assuming a single-step cascade, in which the observed photon energy only arises from one transition. If the available photon energies in $\phi_0(E)$ are much higher than the NRF lines of interest, such that a level of interest $r$ can be reached not just via excitation from the ground state but \textit{also} as the intermediate state during the decay of yet higher excited states (known as `feeding'), more complicated cascades can arise, and a careful sum over Eq.~\ref{eq:triple_diff} must be computed. The 2.212~MeV level of Al-27 is a notable example: it may be reached as an excitation from the ground state or as an intermediate state from the decay of at least 46 different states with level energy $E_r > 2.7$~MeV~\cite{nndc2011al27}. In this work, the photon beam energies are ${\leq}2.7$~MeV, so the strength of the $2.212$~MeV line is driven entirely by the transition from the ground state to the $2.212$~MeV level followed by decay back to the ground state.} in the infinitesimal solid angle $d\Omega$,
\begin{align}\label{eq:triple_diff}
\frac{d^3n}{dE\, d\Omega\, dx} = \phi_t(E) \,b_{r,j}\, \frac{d\mu_\mathrm{NRF}(E)}{d\Omega} e^{-a(x,\theta, E, E')} \epsilon_\mathrm{int}(E') P_f(E')
\end{align}
where $b_{r,j}$ is the branching ratio for the line of interest, $e^{-a(x,\theta,E,E')}$ is a function (described below) that accounts for both the resonant and non-resonant attenuation in the foil, $\theta \equiv \pi - \chi$ is the angle of photon emission relative to the backwards beam direction, $P_f(E')$ is the probability that a photon of energy $E'$ emitted from the foil is transmitted through any intervening filter material designed to reduce count rates at low energies, and $\epsilon_\mathrm{int}(E')$ is the intrinsic photopeak detector efficiency. The angular differential NRF cross section is defined in terms of the angle-integrated cross section as
\begin{align}
\frac{d\sigma_r^\mathrm{NRF}(E)}{d\Omega} = \frac{W(\chi)}{4\pi} \sigma_r^\mathrm{NRF}(E)
\end{align}
where the angular correlation function $W(\chi)$ is symmetric about $\chi = \pi/2$ so that $W(\chi) = W(\theta)$, where again the emission angle $\theta$ is relative to the backwards beam direction. The angular correlation function $W(\theta)$ then takes the form~\cite{hamilton90correlation}
\begin{align}\label{eq:W}
W(\theta) = \frac{1 + a\cos^2\theta + b\cos^4\theta}{1+a/3 +b/5},
\end{align}
which is normalized to $4\pi$ over all solid angles. The constants $a$ and $b$ denote the contribution from dipole and quadrupole transitions, respectively, and are determined by the sequence of spins $J_0 \to J_r \to J_j$. For example, the $0\to 1 \to 0$ spin sequence for the 2.176~MeV transition in U-238 has $a=1$ and $b=0$, and thus $W(\theta) \simeq 1$ at the angle $\theta_d \simeq 55^\circ$ commonly used in experiments. More comprehensive tables of $a$ and $b$ for various spin sequences are given in Ref.~\cite{hamilton90correlation}, where they are denoted by $(R/Q)$ and $(S/Q)$. 
The attenuation factor $e^{-a(x,\theta,E,E')}$ can be decomposed into the attenuation of photons travelling a distance $x$ into the foil (reducing the flux available for NRF interactions) and the attenuation of NRF photons exiting at angle $\theta$ (see Fig.~\ref{fig:schematic}):
\begin{align}
a(x,\theta, E, E') &= a_\mathrm{in}(x,E) + a_\mathrm{out}(x,\theta, E')\\
&= \left[ \mu_\mathrm{NRF}(E) + \mu_\mathrm{nr}(E) \right] x + \left[ \mu_\mathrm{nr}(E') \right] \frac{x}{\cos\theta}.
\end{align}
Altogether, the triple-differential rate is
\begin{align}\label{eq:d3ndEdOmegadx}
\frac{d^3n}{dE\, d\Omega\, dx} = \phi_t(E)\, b_{r,j}\, \mu_\mathrm{NRF}(E) \frac{W(\theta)}{4\pi} \exp\left\{ -x \left[ \mu_\mathrm{NRF}(E) + \mu_\mathrm{nr}(E) + \frac{\mu_\mathrm{nr}(E')}{\cos\theta} \right] \right\} \epsilon_\mathrm{int}(E') P_f(E').
\end{align}
Writing the various $\mu$ terms as
\begin{align}
\mu_\text{eff}(E,E',\theta) \equiv \mu_\mathrm{NRF}(E) + \mu_\mathrm{nr}(E) + \frac{\mu_\mathrm{nr}(E')}{\cos\theta}
\end{align}
for brevity, and integrating over the thickness of the foil from $x=0$ to $x=X$, we have
\begin{align}\label{eq:d2ndEdOmega}
\frac{d^2n}{dE\, d\Omega} = \phi_t(E) \,b_{r,j}\, \mu_\mathrm{NRF}(E) \frac{W(\theta)}{4\pi} \frac{1-\exp\left[ -X \mu_\text{eff}(E,E',\theta) \right] }{\mu_\text{eff}(E,E',\theta)} \epsilon_\mathrm{int}(E') P_f(E'),
\end{align}
which is the most widely-applicable equation for NRF count rates in simple geometries (aside from some possible approximations, special cases, and generalizations, as discussed below). In the general case, since the integrations over $E$ and (possibly) $\Omega$ still need to be carried out numerically, and the $\phi_0(E)$ term may need to be computed using a Monte Carlo simulation, we henceforth refer to Eq.~\ref{eq:d2ndEdOmega} (and its variants) as the \textit{semi-analytical model} for NRF count rates in a transmission measurement with slab components.

In a real HPGe detection system, the NRF flux for a given transition is detected as an approximately Gaussian peak centered at $E'$ with standard deviation $\sigma \sim1$~keV rather than the Doppler width $\Delta \sim 1$~eV of Eq.~\ref{eq:DopplerDelta} due to the effects of detector resolution. This Gaussian NRF peak (or series thereof) sits atop an approximately exponentially-decaying continuum background generated predominantly by secondary electron bremsstrahlung processes in the foil.

\subsection{Considerations for high-accuracy calculations}\label{sec:considerations}
In this section, we discuss additional factors that one must consider in order to perform high-accuracy NRF verification calculations and simulations. These factors are known in the literature, but generally have not been addressed at the level of detail required for the ${\sim}1\%$ absolute agreement between simulation and semi-analytical modeling sought in this work.
\begin{enumerate}
\item \textbf{NRF cross section approximations:}\label{item:xsec_approx}
The primary NRF model approximation found in the literature (e.g., Ref.~\cite{metzger1959resonance}, and the original version of G4NRF) involves an approximation to the NRF cross section equation: rather than exactly compute the Doppler-broadened Breit-Wigner distribution of Eq.~\ref{eq:sigmaNRF}, which involves a numerical integral, various approximations can be made to improve computational performance at the expense of accuracy. One useful and often accurate approximation is a Gaussian NRF cross section, which is achieved by eliminating the $y$ term in the exponential in Eq.~\ref{eq:sigmaNRF} and approximating $t \gg 1$. The result will be inaccurate for large $y$---the non-Gaussian Eq.~\ref{eq:sigmaNRF} retains its Lorentzian tails, which fall off much more slowly than the Gaussian---but the large-$y$ parts of the integrand contribute only a small amount to the integral in Eq.~\ref{eq:d2ndEdOmega}. Changing variables back from $x$ to $E$ we then have
\begin{align}\label{eq:sigmaNRFgaus}
\sigma_r^\text{Gaus}(E) = 2 \pi^{3/2} g_r \left( \frac{\hbar c}{E_r} \right)^2 b_{0,r} \frac{\Gamma_r}{\Delta} \exp\left[ - \frac{(E-E_r)^2}{\Delta^2} \right].
\end{align}
Since Eq.~\ref{eq:sigmaNRFgaus} requires $t \gg 1$ in order to be valid, it is inaccurate for resonances in, e.g.,~Pb, which have relatively large level widths $\Gamma_r$. Moreover, as we will show below, it becomes inaccurate for thick target geometries.

A rougher approximation is a rectangular cross section, i.e., a cross section of constant value within a narrow range of energies that preserves the integrated cross section (Eq.~\ref{eq:sigma_int}), which allows an analytical solution to the integral over $E$ in certain cases~\cite{quiter2010thesis}. While count rates computed with the Gaussian approximation may be accurate to within a few percent (and much faster to evaluate), rates computed using the rectangular approximation may be significantly inaccurate. In Fig.~\ref{fig:sigmadependenceX}, we show---as a function of foil areal density $\rho X$, for two NRF lines---the ratio of count rates predicted by the Gaussian and rectangular cross sections to the baseline prediction from the numerically integrated cross section. All approximations perform worse as $\rho X$ increases, then saturate in inaccuracy at large $\rho X$ as adding more material to the back of the foil only marginally increases the count rate. Similarly, Fig.~\ref{fig:sigmadependenceD} shows the same calculation plotted as a function of object areal density $\rho D$, where no saturation is observed since adding more material to the back of the object will always decrease the transmitted flux. Based on Figs.~\ref{fig:sigmadependenceX} and \ref{fig:sigmadependenceD}, the rectangular cross section approximation should be avoided unless the object and foil are both extremely thin (${\lesssim}1$~g/cm$^2$). The Gaussian approximation is appropriate in intermediate cases where the object is moderately thin (${\lesssim}10$~g/cm$^2$) regardless of the foil thickness. The full numerical integration must be performed to obtain good accuracy in all other scenarios. A similar conclusion was reached in Ref.~\cite[\S3.2.3]{quiter2010thesis}, which considered cross-section-induced errors in the inferred attenuation of an assay target rather than in the absolute count rate. In the updated version of G4NRF, the numerical integration is performed using Simpson's rule, with a user-specified number of meshpoints determining the accuracy of integration.

\begin{figure}[!h]
\centering
\includegraphics[width=0.85\columnwidth]{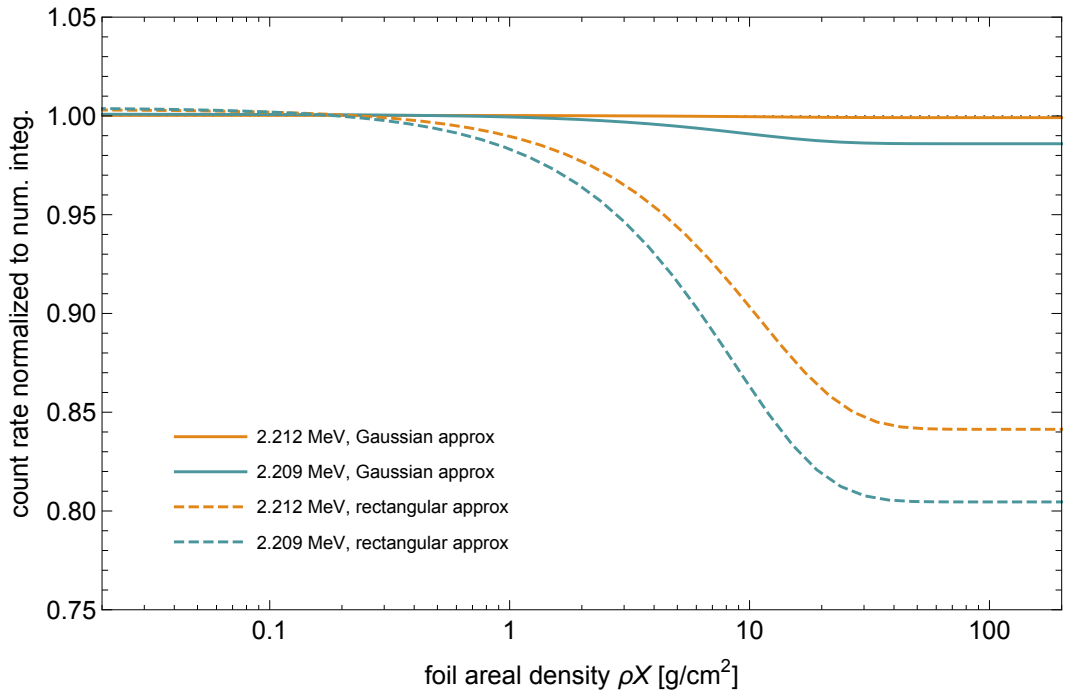}
\caption{Count rates relative to the full numerical integration under approximations of the NRF cross section, as a function of foil areal density $\rho X$. The four curves correspond to the Gaussian approximation (solid curves) for the Al-27 2.212 MeV and U-238 2.209 MeV cross sections and their rectangular analogues (dashed curves), each with $D=0$. The rectangular approximation results in up to $20\%$ error for large foil areal densities~$\rho X$, while the Gaussian approximation results in ${\lesssim}2\%$ error.}
\label{fig:sigmadependenceX}
\end{figure}

\begin{figure}[!h]
\centering
\includegraphics[width=0.85\columnwidth]{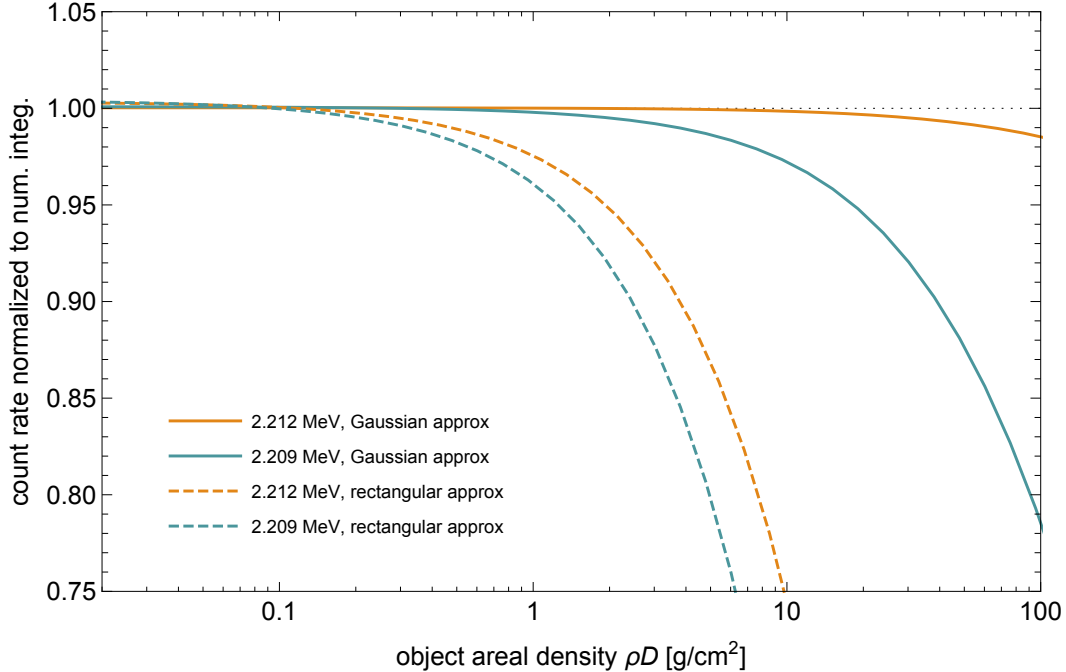}
\caption{Count rates relative to the full numerical integration under approximations of the NRF cross section, as a function of measurement object areal density $\rho D$. The four curves correspond to the Gaussian approximation (solid curves) for the Al-27 2.212~MeV and U-238 2.209~MeV cross sections and their rectangular analogues (dashed curves), each in the limit $X \to 0$. The accuracy of the rectangular approximation is poor even for modest values of $\rho D$, and eventually reaches a value of zero. The Gaussian approximation retains good accuracy over a larger range of $\rho D$, but eventually decays for very thick targets.}
\label{fig:sigmadependenceD}
\end{figure}

\item \textbf{Integration over $\Omega$:} In an idealized experiment in which a relatively small detector is placed far from the foil, we can approximate the integral over $\Omega$, since the $W(\theta)$ and $\cos\theta$ terms can be treated as very nearly constant over the solid angle $\Omega_d$ subtended by the detector:
\begin{align}
\frac{dn}{dE} = \phi_t(E) \,b_{r,j}\, \mu_\mathrm{NRF}(E) W(\theta_d) \frac{\Omega_d}{4\pi} \frac{1-\exp\left[ -X \mu_\text{eff}(E,E',\theta_d) \right] }{\mu_\text{eff}(E,E',\theta_d)} \epsilon_\mathrm{int}(E') P_f(E'),
\end{align}
where $\theta_d$ is the central $\theta$ of the detector. In a more realistic experimental setup, detectors may be placed relatively close to the foil in order to maximize signal; here the $W(\theta)$ and $\cos\theta$ terms still may be approximately constant over the detector, but the acceptance $\Omega_d/4\pi$ should be computed via Monte Carlo solid angle integration for best accuracy. In the verification tests of Section~\ref{sec:verification}, the solid angle $\Omega$ will be quite large ($0\leq\theta\leq\pi/4$), and a full numerical integration must be performed.

\item \textbf{Integration over $E$ and $y$:}\label{item:intEy}
There are effectively two integrals over energy in Eq.~\ref{eq:d2ndEdOmega}: one over the incident photon energies $E$, and one over $y$ to account for the Doppler shift in energy in Eq.~\ref{eq:sigmaNRF}. Due to the complicated $E$-dependence of the integrand in Eq.~\ref{eq:d2ndEdOmega}, there are no useful analytical approximations to the integral over $E$. However, the $\mu_\mathrm{NRF}(E)$ term in front becomes quite small away from $E_r$, and thus integrating over a narrow symmetric interval around $E_r$ (instead of between $0$ and the bremsstrahlung endpoint $E_\text{max}$) may be preferred for reasons of numerical stability and performance. In the semi-analytical calculation, fixed boundaries of $E = E_r \pm 20$~eV are used; the $\pm 20$~eV cutoff is somewhat arbitrary, but is appropriate as it is much larger than the thermal width $\Delta$. In simulation, the integral over $E$ is not performed explicitly, as photon energies are randomly sampled from a given $\phi_0(E)$ distribution.

In G4NRF, integration over $y$ to account for Doppler broadening of the NRF cross section can be performed ``on-the-fly'' (i.e., explicitly integrating Eq.~\ref{eq:sigmaNRF} for every particle) or by table interpolation (i.e., integrating Eq.~\ref{eq:sigmaNRF} for a finely-spaced mesh of values of $E$ at initialization and interpolating between the tabulated values during runtime). Since it avoids performing redundant numerical integrations at every simulation step, the table interpolation technique is computationally more efficient than the on-the-fly evaluation, and the initial cross section table can be computed to higher accuracy than the on-the-fly cross sections without sacrificing overall simulation performance. By default, G4NRF builds a cross section table of $10^4$ points evenly spaced in $E$ between $E_r \pm 10 \Delta$ at $T=300$~K; for each $E$, G4NRF then uses Simpson's rule with $10^4$ meshpoints to integrate Eq.~\ref{eq:sigmaNRF} between $\pm 10\sqrt{2t}$, where $\sqrt{2t}$ corresponds roughly to the `width' of the integrand. To estimate the integration accuracy when using these default parameters, we can make comparisons against a slower but ultra-high-accuracy integration using bounds of $\pm 100\sqrt{2t}$ and $10^6$ meshpoints. The default values produce cross section accuracies of $\mathcal{O}(10^{-14})$ and $\mathcal{O}(10^{-9})$ for photon energies directly on and $3\Delta$ above the $2.176$~MeV resonance, respectively; for the $2.212$~MeV resonance, the accuracies are $\mathcal{O}(10^{-6})$ at both energies. If Eq.~\ref{eq:sigmaNRF} is instead evaluated on-the-fly, the default values are reduced to $300$ meshpoints and $\pm 4\sqrt{2t}$ in order to avoid drastic performance penalties. In the simulations of Section~\ref{sec:verification}, such reduced values would produce count rate results accurate to ${\sim}3\%$ (compared to the table interpolation) in thick U-238 targets and only ${\sim}10\%$ in Al-27, where due to its larger ratio of widths $t$, approximately 1500 meshpoints are required for ${\sim}3\%$ agreement. Finally, in the semi-analytical model, much wider limits of $\pm 50\sqrt{2t}$ can be used without introducing numerical stability issues in the default integration routines of Wolfram Mathematica~\cite{mathematica}.

\item \textbf{Thin target limits:} For thin foils ($X \ll 1/\mu_\text{eff}(E,E',\theta)$), a first-order Taylor expansion of the exponential term collapses Eq.~\ref{eq:d2ndEdOmega} to
\begin{align}
\frac{d^2n}{dE\, d\Omega} \simeq \phi_t(E) \,b_{r,j}\, \mu_\mathrm{NRF}(E) \frac{W(\theta)}{4\pi} X \epsilon_\mathrm{int}(E') P_f(E'),
\end{align}
which is linear in the foil thickness $X$. Moreover, if the measurement object is also thin ($D \ll 1/[\mu_\text{NRF}(E)+\mu_\text{nr}(E)]$), then $\phi_t(E)$ can be approximated as nearly constant over the resonance and taken outside of the $E$ integral. In this case, only the integrated NRF cross section (Eq.~\ref{eq:sigma_int}) is important, rather than the detailed shape of the cross section:
\begin{align}\label{eq:dndOmega_thin}
\frac{dn}{d\Omega} \simeq \phi_t(E_r) \,b_{r,j}\, N \left[\int \sigma_r^\mathrm{NRF}(E)\, dE\right] \cdot \frac{W(\theta)}{4\pi} X \epsilon_\mathrm{int}(E') P_f(E').
\end{align}
Since the rectangular approximation preserves the integrated cross section, Eq.~\ref{eq:dndOmega_thin} explains why the rectangular approximation is accurate for thin targets in Figs.~\ref{fig:sigmadependenceX} and \ref{fig:sigmadependenceD}.

\item \textbf{Notch refill:}\label{item:notch_refill}
Notch refill is the phenomenon in which higher-energy photons are downscattered to the resonance energy $E_r$---typically through small-angle Compton scattering or photoelectron-induced bremsstrahlung---thus refilling the notch created by the $\mu_\text{NRF}(E)$ term in Eq.~\ref{eq:phi_t} and increasing the transmitted flux $\phi_t(E)$ near $E_r$ available to produce NRF in the foil. The effect of notch refill is typically only significant for relatively thick measurement objects (e.g., areal densities of $\rho D \gtrsim 50$~g/cm$^2$~\cite{quiter2011transmission}), which provide many opportunities for downscatter, especially in combination with broad energy spectrum photon beams that have high intensities above $E_r$. Notch refill can also be enhanced by geometric parameters; spatially-broad beams and short path lengths between the measurement object and foil both increase the range of possible angles available for Compton scattering into the notch~\cite{pruet2006detecting}. However, as the processes that cause notch refill are included in the standard Geant4 electromagnetic physics processes, the notch refill contribution to the simulated NRF signal is not directly relevant to the verification tests of G4NRF and thus is outside the scope of this paper.

Correspondingly, Eq.~\ref{eq:d2ndEdOmega} considers only the unscattered component of the transmitted flux $\phi_t(E)$, and thus neglects notch refill. The simplified test simulations of Section~\ref{sec:targets_simple}, however, will not have any significant notch refill component: the simulated DU measurement objects have at most $\rho D \simeq 20$~g/cm$^2$, the simulated Al objects have at most $\rho D \simeq 27$~g/cm$^2$, and the incident beam flux $\phi_0(E)$ consists of narrow intervals around the resonance energies. The simulated heterogeneous measurement object in Section~\ref{sec:verification}, conversely, has a total on-axis areal density of ${\sim}50$~g/cm$^2$ and uses a broad-spectrum bremsstrahlung beam, and thus will exhibit a significant contribution from notch refill. To compare these simulations to the semi-analytical model, we will suppress the simulated notch refill component through a post-processing correction by tallying only the NRF photons that had initial energies inside the $10$~eV-wide energy bin of the sampling distribution (see item~\ref{item:importance_sampling}) containing the corresponding $E_r$. This correction is a simple but useful proxy for a more rigorous check on the photon's scattering history, and gives near-perfect discrimination against notch refill photons up to the negligible fraction (very roughly estimated as $\mathcal{O}(10^{-4})$) of notch refill from initial photon energies that are slightly off-resonance but still within the $10$~eV-wide bin. A yet-simpler possible cut, which does not depend on the energy bin structure and is accurate to approximately $\mathcal{O}(0.5\%)$, is to tally only photons within $\pm 1$~keV of their initial energy. In Section~\ref{sec:verification} we apply the former correction, and henceforth will refer to the corrected spectra as `notch-refill-subtracted' spectra.

\item \textbf{Heterogeneous measurement object:} Eq.~\ref{eq:d2ndEdOmega} can be readily adapted to the case of a measurement object that is heterogeneous along the beam axis. The multi-layer version of the transmitted flux becomes
\begin{align}\label{eq:phi_t_multilayer}
\phi_t(E)=\phi_0(E) \exp\left(-\sum_j D_j [\mu_{\mathrm{NRF},j}(E) + \mu_{\mathrm{nr},j}(E)]\right),
\end{align}
where the sum is over the layers of materials $j$ in the measurement object. If the material undergoing NRF is not of a single isotope, the $\mu_{\mathrm{NRF},j}(E)$ term must be multiplied by the non-unity mole fraction of the NRF isotope.

\item \textbf{Homogeneous multi-isotope foil:} In simulations of multi-isotope foils (such as the geometries of Ref.~\cite{kemp2016physical} we replicate in Section~\ref{sec:targets_complex}), it is convenient to homogenize the foil so that the result does not depend on the ordering of foil layers. The effect of the homogenized foil on the semi-analytical model requires more care. First, the effective attenuation coefficient in the foil is expanded into a sum over the isotopes $i$ in the foil:
\begin{align}\label{eq:mu_eff_homog}
\mu_\text{eff}(E,E',\theta) = N_\text{NRF}\, \sigma_r^\text{NRF}(E) + \sum_i N_i \left( \sigma_{\text{nr},i}(E) + \frac{\sigma_{\text{nr},i}(E')}{\cos\theta} \right).
\end{align}
Second, the number density $N_\text{NRF}$ of the isotope of interest needs to be modified from the `natural' value $N_\text{NRF} = \rho N_\text{Av}/A$ that would be used in the single-isotope Eq.~\ref{eq:d2ndEdOmega}, in order to reflect its homogenization with other isotopes. Moreover, the number density of the isotope may differ between the measurement object and the reference foil if, e.g., materials of different isotopic enrichment are used.

\item \textbf{Importance sampling:}\label{item:importance_sampling}
Because the range of energies likely to trigger NRF ($\Delta \sim 1$~eV) is often small compared to the energy range of realistic photon beams (especially bremsstrahlung), an unweighted simulation that directly samples $\phi_0(E)$ may spend a large fraction of its time simulating off-resonance photons that do not contribute meaningfully to the NRF signal. Running a realistic, broad-spectrum, thick-target simulation until low statistical uncertainty in the NRF peaks is achieved can therefore be prohibitively computationally expensive without importance sampling, i.e., the preferential sampling of $\phi_0(E)$ around the resonance energies of interest $E_r$. To correct for the preferential sampling (through some user-specified distribution in energy $s(E)$), each photon history is given a weight depending on its initial energy $w(E) = \phi_0(E)/s(E)$ (where both $\phi_0(E)$ and $s(E)$ have the same normalization), such that oversampled photons have small weight values. The detected energy spectrum is then computed as a weighted histogram with bin contents given by the sum of weights in each energy bin.\footnote{If the range of weights spans more than about five orders of magnitude, using floating point arithmetic to sum bin contents may lead to substantial roundoff errors. In the ROOT~\cite{brun1997root} framework, for instance, weighted histograms produced by the commonly-used {\tt TTree::Draw()} method by default use floating point arithmetic rather than double precision. The weighted simulation results of Section~\ref{sec:targets_complex} will be inaccurate unless the user ensures the histogram bin contents are {\tt double} instead of {\tt float}, i.e., the ROOT histogram class {\tt TH1D} is used instead of the default {\tt TH1F}.}

\item \textbf{Uncertainties:}\label{item:uncertainties}
In unweighted Monte Carlo simulations of events described by Poisson statistics (e.g., number of NRF photons detected), the relative uncertainty on the number of events asymptotically decays as the inverse square root of the number of events. In weighted simulations (item~\ref{item:importance_sampling}), statistical uncertainties on each bin are computed as the square root of the sum of the squared weights. The statistical uncertainties on NRF peaks in importance-sampled simulations are therefore the sum of many low-weight (over-sampled) photons sampled near the resonance and relatively fewer high-weight (under-sampled) photons sampled above the resonance but downscattered to the resonance through notch refill (item~\ref{item:notch_refill}). In Section~\ref{sec:targets_complex}, quantitative comparisons between the semi-analytical model and simulation will be performed explicitly only for the notch-refill-subtracted spectrum. Unsubtracted results are however included in the later Fig.~\ref{fig:G4validation_BlackSea} for completeness, where it is evident that the under-sampled notch refill photons produce larger statistical uncertainties due to the higher weights. Uncertainties associated with the notch refill subtraction accuracy (as discussed in item~\ref{item:notch_refill}) will be much smaller than the statistical uncertainties obtained in Section~\ref{sec:targets_complex} and are not considered further. Systematic numerical uncertainties associated with numerical integration of the NRF cross section (Eq.~\ref{eq:sigmaNRF}) and transmission model (Eq.~\ref{eq:d2ndEdOmega}) exist, but are treated as known sources of inaccuracy in item~\ref{item:intEy}.

\item \textbf{Computational performance:} As per the discussion on uncertainties (item~\ref{item:uncertainties}), the relative statistical uncertainty in a Monte Carlo NRF simulation decays sub-linearly with the number of events simulated while the computational expense grows linearly. It can therefore be highly desirable to employ both code optimization and variance reduction techniques rather than rely on sheer computing power in order to obtain high precision in a reasonable time. In terms of code optimizations, two important upgrades to the G4NRF source code involve the NRF cross section parameter lookup and calculation. The initial version of G4NRF inefficiently built a string name of the database file every time an NRF cross section was calculated, resulting in several slow string operations per step. We have replaced this with a vector of database names built at initialization that is accessed during the NRF cross section calculation, resulting in a ${\sim}20\%$ event rate increase when using the on-the-fly cross section calculation of Eq.~\ref{eq:sigmaNRF}.

As discussed in items~\ref{item:xsec_approx} and~\ref{item:intEy}, G4NRF performs numerical integration of the NRF cross section (Eq.~\ref{eq:sigmaNRF}) in order to maintain high accuracy in predicted count rates. The numerical integration in Eq.~\ref{eq:sigmaNRF} is necessarily more computationally expensive than the Gaussian approximation of Eq.~\ref{eq:sigmaNRFgaus}, and thus (if evaluated on-the-fly) trades off statistical confidence in a fixed runtime for reduced systematic error in the cross section evaluation. The accuracy of integration (through Simpson's rule, see item~\ref{item:intEy}) is determined by the number of meshpoints in the integrand, which, when evaluated on-the-fly, the user can adjust to balance evaluation time and accuracy. In the table interpolation method, the user can balance the lower computation per event against the increased overhead of building the tables, which may be expensive if many NRF lines are available. In one simulation, the table interpolation improved event rates by ${\sim}40\%$ (and could likely be optimized even further) at the expense of only $2$~cpu-seconds per cross section table built at initialization. For an older but more comprehensive list of updates to the G4NRF source code, including non-performance-oriented upgrades, the reader is referred to Ref.~\cite{vavrek2016simulations}.

Significant performance gains can also be made in the user-supplied Geant4 code, especially by the variance reduction methods of killing and importance sampling. Killing secondary electrons produced by photon scattering in the measurement object, which are computationally expensive to track but do not influence the detectable NRF signal (though may produce background and notch refill), can result in a ${\sim}25\%$ performance improvement. More consequentially, importance sampling of the incident photon beam around the NRF resonances of interest (item~\ref{item:importance_sampling}) can improve the efficiency of simulation (and thus statistical confidence) in the NRF lines by orders of magnitude, at the price of decreased confidence in the rest of the energy spectrum.

\end{enumerate}

\section{Verification simulations and results}\label{sec:verification}
Upon its initial release, the G4NRF Monte Carlo code was verified (or `benchmarked') against theoretical predictions to a level of ${\sim}20\%$~\cite{jordan2007simulation}. This comparison was made, however, using a considerably simplified semi-analytical model that neglected important effects such as non-resonant attenuation of NRF photons, the non-isotropic nature of NRF emission, and the inaccuracy of the Gaussian cross section approximation (Eq.~\ref{eq:sigmaNRFgaus}) for large resonance widths or thick targets. In this article, we greatly improve the verification of G4NRF to a level of $1$--$3\%$.\footnote{In a preliminary work, we achieved a $5$--$10\%$ agreement~\cite{vavrek2018transport}.} Such agreement is found by performing a series of straightforward but powerful tests based on Eq.~\ref{eq:d2ndEdOmega}, a semi-analytical model that does not make the three aforementioned approximations. In particular, we first use Eq.~\ref{eq:d2ndEdOmega} and the assumed data in Table~\ref{tab:xsec_params} to compute the prediction for the NRF count rates in each of the three most prominent U-238 resonances and the strong Al-27 resonance near $2.2$~MeV in simple homogeneous (isotopically pure) geometries and directly compare these results against Geant4+G4NRF simulations of the same test cases.\footnote{Simulation and semi-analytical calculation codes (written in C++ and Mathematica, respectively) are available at {\tt https://github.com/jvavrek}.} We then consider a more complex scenario---the NRF measurement first simulated in Ref.~\cite{kemp2016physical}---involving isotopically heterogeneous geometries, a simulated bremsstrahlung beam, and non-isotropic NRF emission, and compare NRF count rates in two lines of U-238 and one of Pu-240.

\subsection{Homogeneous targets}\label{sec:targets_simple}
The test flux $\phi_0(E)$ in each homogeneous target simulation is composed of $10^9$ photons with energies uniformly random sampled within $\pm 25$~eV of the resonance energies $E_r$ of interest. A thin pencil beam of these photons is directed orthogonally at an isotopically-pure rectangular slab object with transverse dimensions of $30 \times 30$~cm and a linear thickness $D$ of $0$, $0.1$, $1$, or $10$~cm. In the slab, the photons undergo both resonant (NRF) and non-resonant scattering. The flux transmitted through the slab then impinges on the reference foil (placed $100$~cm downstream), which is another isotopically-pure rectangular slab with transverse dimensions of $50 \times 50$~cm and a linear thickness $X$ ranging from $0.02$--$20$~cm. Again, photons interact both resonantly and non-resonantly in the foil, and the energies $E'$ of all photons exiting the foil with polar angles $\theta \leq \pi/4$ relative to the backwards beam direction are recorded (essentially setting $\epsilon_\mathrm{int}=1$ and $P_f = 1$). To further simplify the comparison, the simulated NRF photon emission is set to be isotropic, and $W(\theta)$ is correspondingly set to $1$ in the semi-analytical model.

\begin{figure}[!h]
\centering
\includegraphics[width=0.85\columnwidth]{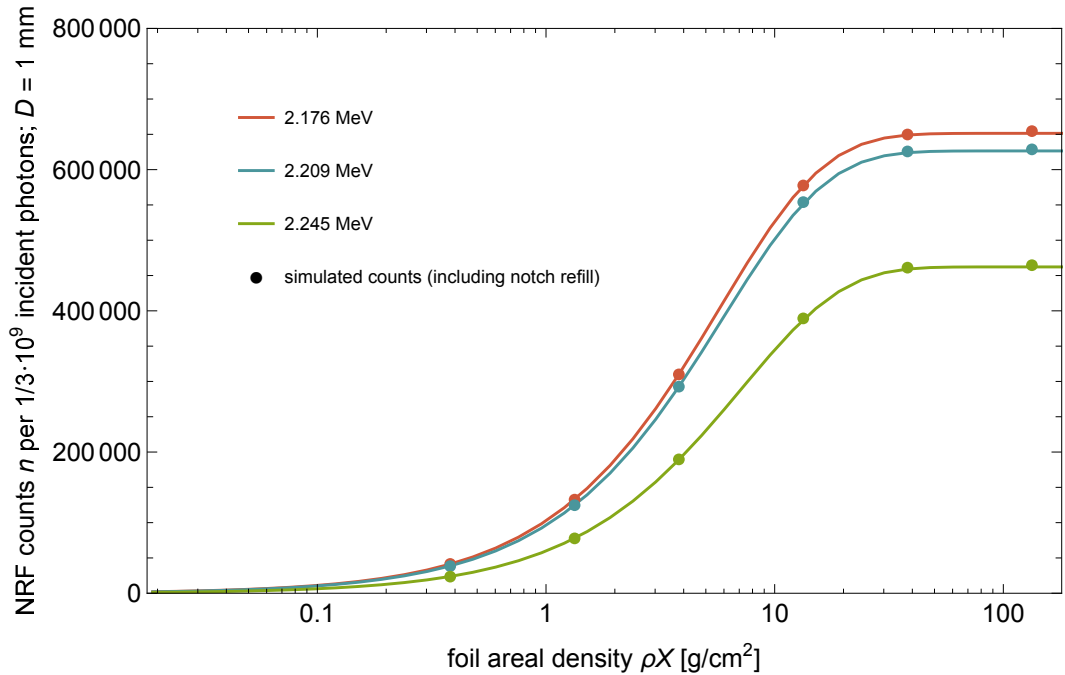}
\caption{Verification of G4NRF in U-238 with a thin ($D=1$~mm) target. Solid lines correspond to the semi-analytical model predictions (via integration of Eq.~\ref{eq:d2ndEdOmega}), while points correspond to the count rates produced using G4NRF. Here (and in subsequent Figures) error bars denote $\pm 1\,\sigma$ statistical uncertainties. Error bars are smaller than the size of the markers.}
\label{fig:G4validation_U_D1mm}
\end{figure}

\begin{figure}[!h]
\centering
\includegraphics[width=0.85\columnwidth]{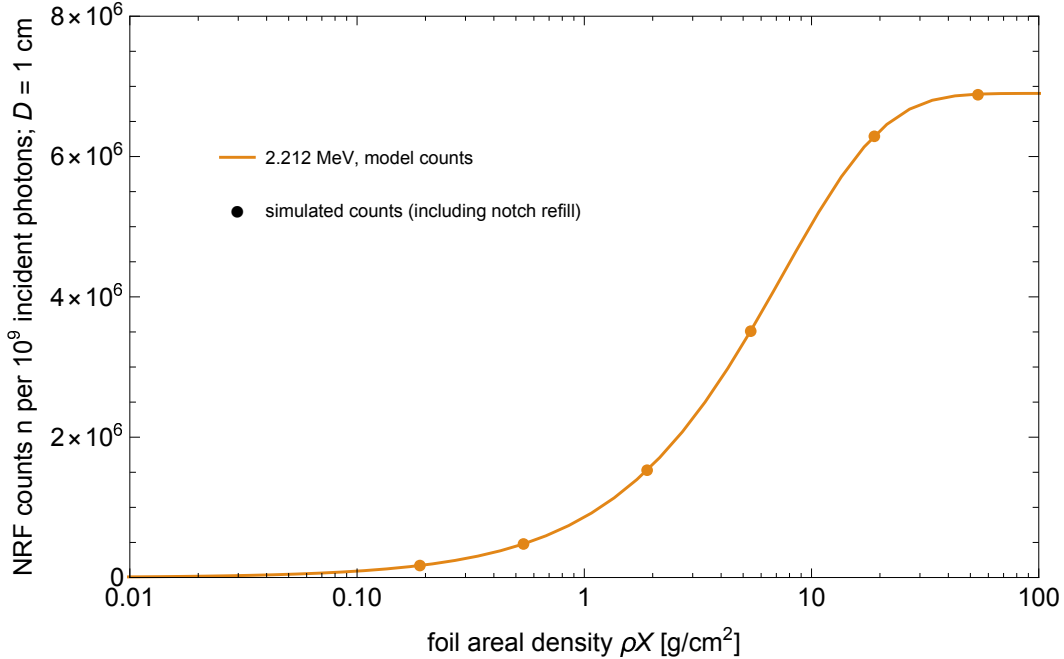}
\caption{Verification of G4NRF in Al-27 with a thin ($D=1$~cm) target.}
\label{fig:G4validation_Al_D1cm}
\end{figure}

Fig.~\ref{fig:G4validation_U_D1mm} compares Eq.~\ref{eq:d2ndEdOmega} to the Monte Carlo results for the representative scenario of a thin U-238 target and a U-238 foil of variable thickness $X$. The maximum relative deviation between the simulation and semi-analytical model (over all $\rho X$) is ${\sim}1\%$ for each of the three U-238 NRF lines. For the thinnest $\rho X$ and thickest $\rho D$ case (i.e., the weakest NRF signal), this ${\sim}1\%$ agreement is on the order of the expected $1.3\%$ statistical fluctuations. Similarly, Fig.~\ref{fig:G4validation_Al_D1cm} shows the comparison for a thin Al-27 target and an Al-27 foil. A full list of maximum relative deviations, all ${\sim}1\%$, is given in Table~\ref{tab:verification_results}. The good agreement in the thick U-238 scenario ($\rho D = 19$~g/cm$^2$) of Table~\ref{tab:verification_results} moreover suggests that a notch refill correction to Eq.~\ref{eq:d2ndEdOmega} is unnecessary at object areal densities of ${\lesssim}20$~g/cm$^2$ when using narrow-energy beams, but is important at higher areal densities with broad-spectrum beams as discussed below. In both U-238 and Al-27, the largest maximum discrepancies listed in Table~\ref{tab:verification_results} occur for the targets with the largest $\rho D$, perhaps suggesting that a small $\mathcal{O}(1\%)$ systematic bias in the NRF cross section evaluation or in the treatment of non-resonant interactions has not yet been accounted for.

\subsection{Heterogeneous targets}\label{sec:targets_complex}
A more stringent verification test (and one tied directly to the nuclear treaty verification application of NRF) is the retrodiction of count rates in the simulations of Ref.~\cite{kemp2016physical}, which use multi-isotope measurement objects and foils, a realistic incident beam of $2.7$~MeV endpoint bremsstrahlung importance-sampled around the resonances of interest, and non-isotropic NRF emission. The treaty verification scenario outlined in Ref.~\cite{kemp2016physical} involves distinguishing (via NRF measurements) potential hoax nuclear warheads from a (proxy) genuine nuclear warhead design based on the `Black Sea' experiments~\cite{fetter1990gamma}. Here we study the `geometric hoax' object of Ref.~\cite{kemp2016physical}, which is a series of rectangular slabs that matches the on-axis sequence of materials of the genuine Black Sea warhead (in a single orientation) and thus could be used in an attempt to cheat simpler types of verification measurements such as radiography. This geometric hoax object therefore consists of three pairs of $35.5 \times 35.5$~cm rectangular slabs of different thicknesses and materials: the two outermost layers are each $0.25$~cm of uranium with an isotopic content of $95/5\; \text{wt}\%$ U-235/238, while the two innermost layers are each $0.43$~cm of plutonium with an isotopic content of $94/6\; \text{wt}\%$ Pu-239/240. Between each set of uranium and plutonium layers sits $6.5$~cm of HMX-type explosive. The reference foil, again $100$~cm downstream from the center of the measurement object and $50 \times 50$~cm in the transverse beam directions, is a homogenized mixture of U-235, U-238, Pu-239, and Pu-240 (all equal parts by mass, with an overall density of $\rho = 19$~g/cm$^3$), varying in linear thickness $X$ from $0.02$--$7$~cm.

\begin{figure}[!h]
\centering
\includegraphics[width=0.85\columnwidth]{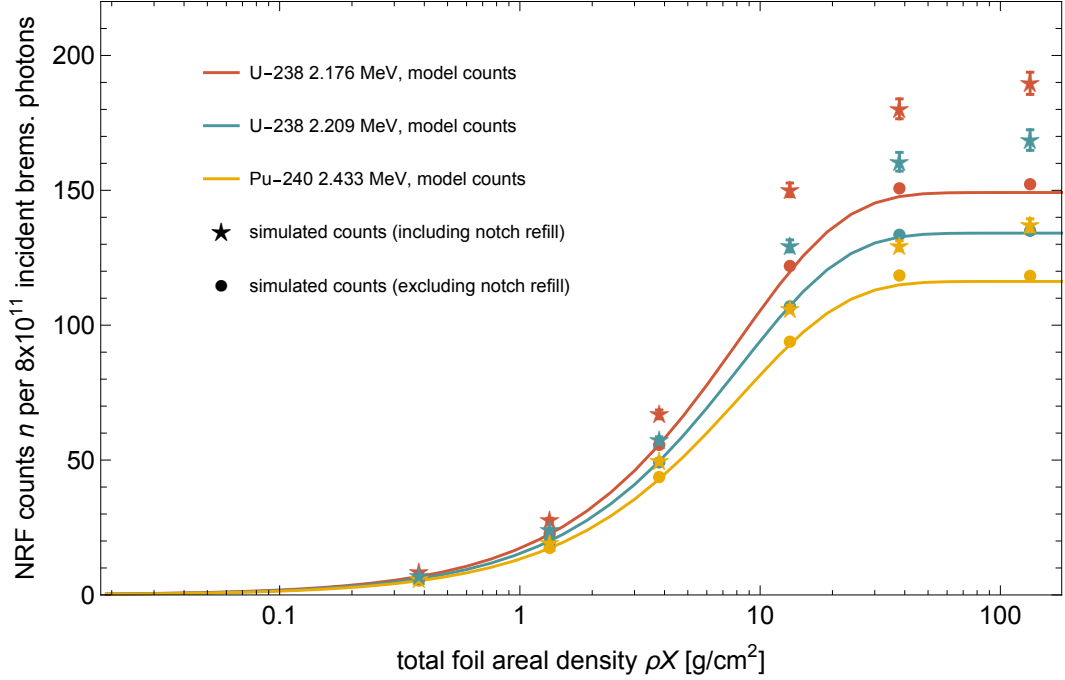}
\caption{Verification of G4NRF in the simulations of Ref.~\cite{kemp2016physical}. Circles denote the notch-refill-subtracted results and stars denote the raw results. Following the convention of Ref.~\cite{kemp2016physical}, values of NRF counts $n$ are given per $8\times 10^{11}$ (importance-sampled) primary bremsstrahlung photons. Statistical uncertainties in the notch-refill-subtracted results are smaller than the size of the points. Uncertainties in the unsubtracted results are larger due to the importance sampling scheme, since the off-resonant parts of $\phi_0$---and thus the photons that contribute to notch refill---are undersampled compared to the resonant energy bins (see Section~\ref{sec:considerations}, items~\ref{item:notch_refill}, \ref{item:importance_sampling}, and \ref{item:uncertainties}).}
\label{fig:G4validation_BlackSea}
\end{figure}

Using the extended semi-analytical model of Eqs.~\ref{eq:d2ndEdOmega}, \ref{eq:phi_t_multilayer}, and~\ref{eq:mu_eff_homog}, we calculate the expected count rates for three NRF lines---two U-238 and one Pu-240---for the `geometric hoax' object geometry described above. The calculation is performed for various foil thicknesses $X$, keeping the relative abundances of the four foil isotopes constant. Simulations are performed using a thin pencil beam, with energies importance sampled around the resonant energies of interest. A statistical uncertainty of ${<}1.2\%$ in the simulated NRF rates is achieved by sampling up to $4.5\times 10^{11}$ total bremsstrahlung photons for the smallest foil thicknesses $X$. Fig.~\ref{fig:G4validation_BlackSea} compares the results of these simulations to the semi-analytical model; the two agree to ${\lesssim}3\%$ (maximum deviation) if notch refill and background photons (which are automatically simulated by the standard Geant4 electromagnetic processes) are excluded by the notch refill subtraction procedure outlined in Section~\ref{sec:considerations}, item~\ref{item:notch_refill}. Raw results without enforcing the notch refill cut are also shown as stars in Fig.~\ref{fig:G4validation_BlackSea}, where it is clear that notch refill becomes a significant effect for this thicker (${\sim}50$~g/cm$^2$) object geometry and broad-spectrum beam, inducing more NRF counts than Eq.~\ref{eq:d2ndEdOmega} alone would predict. It should be emphasized again, however, that the contribution from notch refill is not relevant to the verification of G4NRF itself, as notch refill is generated by standard electromagnetic processes already modeled in Geant4, and only acts to increase the flux $\phi_t(E)$ available for NRF production in the foil. A list of results excluding notch refill is included in Table~\ref{tab:verification_results}. The observed discrepancies continue the systematic trend of Section~\ref{sec:targets_simple}, where larger discrepancies between the model and simulation are found in targets with larger $\rho D$. The heterogeneous target discrepancies are typically a result of the model underpredicting the simulation (see Fig.~\ref{fig:G4validation_BlackSea}), while any systematic trend of over- or underprediction for the homogeneous targets of Section~\ref{sec:targets_simple} is much less clear.

\begin{table}[!h]
\centering
\small
\begin{tabular}{c|c|c|c|c}
\# & object and $\rho D$ & foil and $\rho X$ & $E_r$ [MeV] & min, max, RMS discrep [\%]\\\hline
0 & none & U-238, 0.38--133~g/cm$^2$ & 2.176 & 0.15,\; 0.80,\; 0.37\\
 & & & 2.209 & 0.01,\; 0.42,\; 0.25\\
 & & & 2.245 & 0.05,\; 0.79,\; 0.43\\\hline
1 & U-238 plate, 1.9~g/cm$^2$ & U-238, 0.38--133~g/cm$^2$ & 2.176 & 0.00,\; 0.54,\; 0.32\\
 & & & 2.209 & 0.04,\; 0.51,\; 0.32\\
 & & & 2.245 & 0.28,\; 1.02,\; 0.57\\\hline
2 & U-238 plate, 19~g/cm$^2$ & U-238, 0.38--133~g/cm$^2$ & 2.176 & 0.31,\; 1.36,\; 0.84\\
 & & & 2.209 & 0.01,\; 1.51,\; 0.69\\
 & & & 2.245 & 0.16,\; 1.26,\; 0.91\\\hline
3 & none & Al-27, 0.19--54~g/cm$^2$ & 2.212 & 0.03,\; 0.57,\; 0.24\\\hline
4 & Al-27 plate, 2.7~g/cm$^2$ & Al-27, 0.19--54~g/cm$^2$ & 2.212 & 0.03,\; 0.34,\; 0.17\\\hline
5 & Al-27 plate, 27~g/cm$^2$ & Al-27, 0.19--54~g/cm$^2$ & 2.212 & 0.25,\; 1.62,\; 0.82\\\hline
6 & `geometric hoax' object, 51~g/cm$^2$ & four isotopes, 0.38--133~g/cm$^2$ & 2.176 & 0.10,\; 2.13,\; 1.64 \\
 & & & 2.209 & 0.23,\; 1.95,\; 1.08\\
 & & & 2.433 & 0.62,\; 3.01,\; 1.76\\
\end{tabular}
\caption{Minimum, maximum, and root-mean-square (RMS) discrepancies between Eq.~\ref{eq:d2ndEdOmega} (using Eq.~\ref{eq:phi_t_multilayer} and Eq.~\ref{eq:mu_eff_homog} for the multi-isotope cases) and G4NRF simulations for various verification scenarios. In the last row, the `geometric hoax' object and the four-isotope foil are described in the text and in Ref.~\cite{kemp2016physical}, and only the notch refill-subtracted results are compared.}
\label{tab:verification_results}
\end{table}

\section{Discussion and conclusions}\label{sec:discussion}
We have enhanced the accuracy and performance of the G4NRF Monte Carlo code, greatly improving the verification of its simulation results against theory. The verification tests of Section~\ref{sec:verification} indicate that the Geant4+G4NRF Monte Carlo simulations predict the NRF count rates found via the semi-analytical model (e.g., Eq.~\ref{eq:d2ndEdOmega}) to within average maximum deviations of about $1$--$3\%$ for several resonant energies $E_r$ near $2.2$~MeV---see Table~\ref{tab:verification_results}. Such agreement is found both in simple simulations of single-isotope geometries using idealized photon distributions (Section~\ref{sec:targets_simple}), as well as the more complicated geometries and beam fluxes of Ref.~\cite{kemp2016physical} (Section~\ref{sec:targets_complex}). This suggests that given accurate knowledge of NRF cross section parameters (e.g., the resonance widths $\Gamma_{r}$ in Table~\ref{tab:xsec_params}), Geant4+G4NRF can be used to accurately design and analyze the results of future NRF experiments. Moreover, for experiments involving fairly thick measurement object geometries ($\rho D \gtrsim 50$~g/cm$^2$), Geant4+G4NRF is more accurate than the semi-analytical model due to the automatic simulation of notch refill. Further work on G4NRF will focus on preparing the source code for inclusion into the standard Geant4 distribution. Further verification work may involve expanding the semi-analytical model to the entire cascade of NRF photons produced through multiple intermediate energy levels, or modeling the mixing of angular correlation functions for odd-$A$ nuclei. Comparison of the semi-analytical model and G4NRF against the experimental data of Ref.~\cite{vavrek2018experimental} is covered in a forthcoming paper~\cite{vavrek2018absolute}.

\section*{Acknowledgements}
This work was funded in-part by the Consortium for Verification Technology under Department of Energy National Nuclear Security Administration award number DE-NA0002534. BSH gratefully acknowledges the support of the Stanton Foundation's Nuclear Security Fellowship program. The authors wish to thank the Massachusetts Green High Performance Computing Center (MGHPCC) for computing resources, R.~Scott Kemp for providing the derivation of Eq.~\ref{eq:E_rec}, Cody Dennett for useful discussions on nuclear recoil in solids, and the anonymous referee for helpful comments.

\bibliographystyle{unsrt}
\bibliography{bibliography.bib}

\end{document}